\documentclass[preprint,aps,psfig,amssymb,superscriptaddress]{revtex4}
\usepackage{times,floatflt}
\usepackage{amsmath}
\usepackage{epsfig}
\usepackage[]{graphicx}

\begin{document}
\title{Influence of well width fluctuations on the binding energy of excitons,
charged excitons and biexcitons in GaAs-based quantum wells}




\author{A. V.~Filinov}
\affiliation{Departement Natuurkunde, Universiteit Antwerpen (Drie
Eiken Campus), Universiteitsplein 1, B-2610 Antwerpen, Belgium}
\affiliation{Institute of Spectroscopy RAS, Moscow region,
Troitsk, 142190, Russia}
\affiliation{Christian-Albrechts-Universit\"at zu Kiel, Institut
f\"ur Theoretische Physik und Astrophysik, Leibnizstrasse 15,
24098 Kiel, Germany}
\author{C.~Riva}
\affiliation{Departement Natuurkunde, Universiteit Antwerpen (Drie
Eiken Campus), Universiteitsplein 1, B-2610 Antwerpen, Belgium}
\author{F. M.~Peeters}
\affiliation{Departement Natuurkunde, Universiteit Antwerpen (Drie
Eiken Campus), Universiteitsplein 1, B-2610 Antwerpen, Belgium}
\author{Yu. E.~Lozovik}
\affiliation{Institute of Spectroscopy RAS, Moscow region,
Troitsk, 142190, Russia}
\author{M.~Bonitz}
\affiliation{Christian-Albrechts-Universit\"at zu Kiel, Institut
f\"ur Theoretische Physik und Astrophysik, Leibnizstrasse 15,
24098 Kiel, Germany}

\begin{abstract}
We present a first-principle path integral Monte-Carlo (PIMC)
study of the binding energy of excitons, trions (positively and
negatively charged excitons) and biexcitons bound to single-island
interface defects in quasi-two-dimensional
GaAs/Al$_{x}$Ga$_{1-x}$As quantum wells. We discuss in detail the
dependence of the binding energy on the size of the well width
fluctuations and on the quantum-well width. The numerical results
for the well width dependence of the exciton, trions and biexciton
binding energy are in good quantitative agreement with the
available experimental data.
\end{abstract}

\maketitle                   

\section{INTRODUCTION}\label{intro}

Excitonic atoms and molecules in quantum confined semiconductors
have been intensively investigated in the last decade. These
systems show nontrivial Coulomb correlation effects leading to
interesting optical and transport characteristics not seen in bulk
materials. A strong increase of the binding energy of the
excitonic complexes was found experimentally
\cite{exp_xminus,exp_xplus,Brunner, birkedal, Adachi, Panthke, Bar-Ad, Kim,
Smith, langbein} with decreasing quantum well (QW) width and
increasing magnetic field \cite{review}.

In the literature there has been an active discussion about the
influence of localization potentials on the binding energy of
excitons and excitonic complexes \cite{review}. Most of the
theoretical calculations \cite{Riva1, Riva2, Ess, Tsuch,
Stebe2000,Whitt} show a substantially weaker dependence of the
binding energies on the QW width than those experimentally
observed~\cite{exp_xminus, exp_xplus,birkedal, Adachi, Panthke,
Bar-Ad, Kim, Smith, langbein}. In particular, it was found
experimentally that the binding energy of the charged excitonic
complex is for narrow quantum wells much larger than the one
estimated theoretically. The difference being typically a factor
of two for narrow QWs. An explanation for this could be the
trapping of the excitons (trions, biexcitons) by ionized donors in
the barriers~\cite{Eytan} or by some kind of interface defects
produced by the mixture of well and barrier materials during the
QW growth process, i.e. by QW width fluctuations or fluctuations
in the alloy composition of the barrier, which were not taken into
account in Refs.~\cite{Riva1, Riva2, Ess, Tsuch, Stebe2000,
Whitt}. Such effects can induce an additional weak lateral
confinement which leads to the confinement of the particles in all
three dimensions like in the case of a quantum dot potential. The
low-temperature photoluminescence of such structures originates
from the radiative recombination of the exciton states localized
at such nonuniformities of the heterostructure potential. In this
situation broadening and splittings of both exciton and trion
peaks develops in the PL spectra~\cite{Gam}. Such lateral
confinement becomes more important in narrow quantum wells.
Theoretical calculations of exciton, trion and biexciton states in
such structures is a fairly complex problem because of the need to
take simultaneous account of the Coulomb interaction and the
three-dimensional heterostructure potential, which is no longer
translationally invariant.

The standard theoretical approach to calculate binding energies is
to solve the corresponding many-particle Schr\"odinger equation by
means of an appropriate basis expansion. This works efficiently in
simple geometries but is not easily applicable to our problem with
well-width fluctuations. Recently, a different approach was
developed which is based on solving the Bloch equation for the
many-particle density matrix \cite{png2}. It was demonstrated in
Ref.~\cite{png2} that this problem can be efficiently solved using
path integral Monte Carlo (PIMC) methods without any restrictions
on the geometry of the confinement potential.
No quantum well
width fluctuation effects were considered in Ref.~\cite{png2}.

The aim of the present paper is to understand and explain recent
experimental data on the binding energy of the ground state
excitons, trions, biexcitons in QW's by including localization
effects. We consider localization as a consequence of the local
modulation of the thickness of the quantum well of $1-2$
monolayers which corresponds to the experimental findings of
Ref.~\cite{Gam}.
In agreement with experimental results, we find that such
QW width fluctuations can
increase the trion binding energy in GaAs-based quantum wells by
up to 100 percent as compared to ideal QW's without interface
roughness.

We also found that for lateral localization
diameters exceeding $D \approx
150$ {\AA}
the binding energy of the negative trion can become larger
than that of the positive trion, in contrast to the case of ideal
QW's where the positively charged excitons are slightly more strongly
coupled. The reason is that the localization confinement has a
different influence on the lateral wave functions of electrons and
holes. Thus the trion composition (i.e. $X^+$ versus $X^-$)
becomes crucial to the value of
the binding energy when the localization diameter changes.

Our numerical method is an extension of the Path Integral approach
of Ref.~\cite{png2}. The method does not involve expansions in
terms of basis functions, no symmetry assumptions are made (in
this sense it can be considered as first-principle), and the error
can be managed in a controllable way~\cite{Feynm}.

The paper is organized as follows. In Sec.~\ref{model} we present
and discuss the Hamiltonian for the exciton, biexciton and charged
excitons in a quantum well with interface defects. We also discuss
the approximations used in the present calculations. In
Sec.~\ref{PIMCA} we introduce the basic ideas of the numerical
method, i.e. Path Integral Monte Carlo (PIMC), used to obtain the
ground state of the excitonic complexes. In Sec.~\ref{results1} we
compare the correlation, localization and binding energy of the
localized exciton ($X$), biexciton ($X_2$) and charged excitons
($X^{\pm}$) ground state with the ones of the non localized, i.e.
free exciton complexes in a quantum well. Further, we study the
dependence of the $X$, $X^{\pm}$ and $X_2$ ground state properties
on the defect width and height. In Sec.~\ref{experiment} we
compare our calculations with the available experimental data and,
finally, present our conclusions in Sec.~\ref{conclusion}.

\section{The theoretical model}\label{model}

We consider a single GaAs quantum well grown between two Al$_{x}$Ga$_{1-x}$As barriers.
The effective mass framework is used to describe the semiconductor material and the QW
structure. Using the isotropic approximation the Hamiltonian for $N_e$ electrons and
$N_h$ holes reads:
\begin{eqnarray}
H = \sum\limits_{i=1}^{N_e,N_h} \left[ -\frac{\hbar^2}{2 m_i} \nabla^2 + V_{e\,(h)}(z_i)
+  V^{loc}_{e\,(h)}(\bf{r}_i) \right]
+ \sum\limits_{i<j}^{N_e,N_h} \frac{e_i\, e_j}{\epsilon |{\bf
r}_i- {\bf r}_j|}&,& \label{3D_Ham}
\end{eqnarray}
where $m_i$ and $e_i$ are the mass and charge of the i-th particle, $\epsilon$  is the
dielectric constant, which we assume equal for the well and for the barrier, $V_{e\,(h)}$
is the confinement potential associated with the presence of the QW, $V^{loc}_{e\,(h)}$
is the lateral (localization) confinement which is due to the fluctuations of the QW
width. We take the quantum well growth direction as the $z$-direction.

For a GaAs/Al$_{x}$Ga$_{1-x}$As quantum well, we consider the
following heights of the square-well potential: $ V_e =
0.57\times(1.155x + 0.37x^2)$ $eV$ for electrons and $ V_h =
0.43\times(1.155x + 0.37x^2)$ $eV$ for holes. In our calculations
we use an $Al$ concentration of $x=0.3$. Furthermore, the
following material parameters are used: $\epsilon =12.58$, $m_e =
0.067 \, m_0$, $m_h = 0.34 \, m_0$, where $m_0$ is the mass of the
free electron. The units for energy and distance are  $H_a^*=2
R_y^{*} = e^2 / (\epsilon \, a_B) = 11.58 \; meV$, $a_B = \hbar^2
\epsilon / (m_e \, e^2) = 99.7$ {\AA}, respectively. We have also
considered the case of an anisotropic hole mass according
to~\cite{Winkler}, using for the in-plane hole mass a smaller
value of $m_h^{||}=0.112 m_0$, and in the quantum well growth
direction $m_h^z=0.377 m_0 $. Comparing the binding energies
calculated with the isotropic and anisotropic approximations gives
important insight about the relevance of band structure details
for the properties of excitonic complexes in quantum wells.

\begin{figure}
\includegraphics[width=.7\textwidth,  angle=-90]{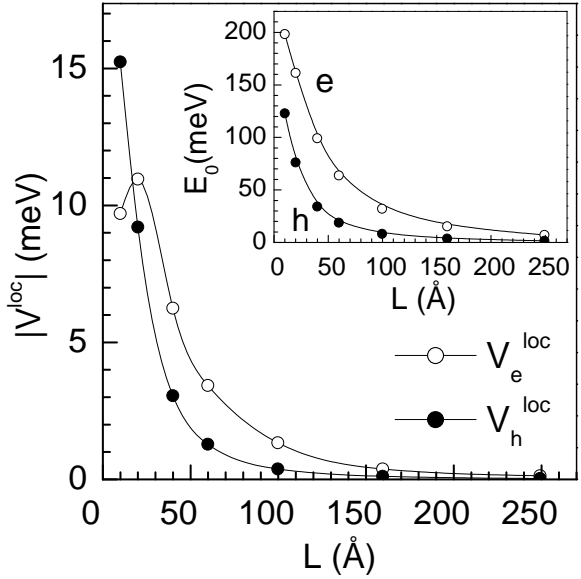}
\vspace{-2.4cm} \caption{Dependence of the height of the
localization potential $V^{loc}_{e(h)}$, Eq.~(\ref{uloc2d}), for
electrons (open dots) and holes (full dots) on the well width $L$
for a well width fluctuation of one monolayer. Inset: the lowest
energy level in the potential, $V_{e(h)}$, vs. the well width.}
\label{u_loc}
\end{figure}

The actual shape of the interface defects is not known and depends
on the sample growth conditions. To limit the number of
parameters, we simulate the interface defects through a
cylindrically symmetric potential with a lateral radius $R$ and
height $V^{loc}_{e,h}$. The potential height is determined by the
zero-point energy and was obtained as the difference between the
lowest energy levels of the electron (hole) in two QW's with the
widths $L$ and $L+\delta$, where $\delta=na$ with $n=1,2$
and $a$ is the thickness of a single monolayer. Because of the
difference in mass between the electron and the hole, the height
of the localization potential will also be different (see
Fig.~\ref{u_loc}). For large well widths (L) the localization
potential is given by $(\hbar^2\pi^2/m_i)\delta/L^3$. Notice that
in Fig.~\ref{u_loc} for small L the electron localization has a
local maximum which is due to the increased penetration of the
electron wave function into the barrier material. For the hole
this occurs at much smaller L due to its larger mass.

In GaAs, 1 (2) monolayer(s) correspond to a well width fluctuation
of $\delta = 2.8$ ($5.6$) {\AA}. These parameters ensure that the
exciton (trion, biexciton) wave function in the growth direction
$z$ is practically not affected by the defect. It is very
instructive to see from Fig.~1 that, for narrow QW's, this lateral
localization potential reaches about $15$ meV which is comparable
to the exciton binding energy and is several times larger than the
trion binding energy. This behavior is in qualitative and
quantitative agreement with the monolayer splitting measured
experimentally~\cite{Gam}. In the inset of Fig.~\ref{u_loc} we
plot the value of the lowest energy level in a QW
as a function of the QW width.
The main figure can be obtained directly from the results of
the inset through $V^{loc} = E_0(L+\delta) - E_0(L)$.

We proceed further with the assumption that the QW confinement is
sufficiently strong and that the Coulomb interaction among the
particles in the $z$-direction will not modify the wave functions
in the $z$-direction, and consequently we may use in the
$z$-direction the noninteracting electron and hole wave functions.
In this {\em adiabatic approximation} we neglect the influence of
in-plane electron-hole correlations on their motion perpendicular
to the QW plane. This assumption is valid due to the strong
quantization in square wells of widths $L \lesssim a_B$, giving
rise to the condition on the energies: $\Delta E^z_{e \, (h)} \gg
E_c, E_b^X, E_b^{X_2}, E_b^{X^{\pm}}$, where $\Delta E^z_{e \,
(h)}$ is the level spacing in the quantum well, and $E_c, E_b^X,
E_b^{X_2}, E_b^{X^{\pm}}$ are the correlation and binding energy
of exciton, biexciton and trions, respectively.

Our approach to compute the binding energies starts from the
$N$-particle ($N=2,3,4$) density matrix of the excitonic complex
of interest (exciton, trion, biexciton) which is obtained from a
solution of the corresponding Bloch equation, see
Ref.~\cite{png2}. In the adiabatic approximation, the full
$N$-particle density matrix factorizes into
\begin{equation}
\rho({\bf R}^{xyz}, \beta) = \rho(Z_e,\beta) \, \rho(Z_h,\beta) \,
\rho({\bf R}^{xy},\beta),
\label{adiab}
\end{equation}
where ${\bf R}^{xyz} \,({\bf R}^{xy})= \{{\bf r}_{e\,1}, {\bf
r}_{e\,2},\dots,{\bf r}_{e \, N_e};\; {\bf r}_{h\,1}, {\bf
r}_{h\,2},\dots,{\bf r}_{h \, N_h}\}$ is a $3D$ ($2D$) vector of
all particle coordinates, $Z_{e\, (h)}$ is the $z$ coordinate of
all electrons (holes), $\rho(Z_e,\beta)$ and  $\rho(Z_h,\beta)$
are the density matrices of free electrons and holes confined in
the $z$ direction by the square well, and $\beta=1/k_B T$ is the
inverse temperature. We underline that the density matrix
$\rho({\bf R}^{xy},\beta)$ contains all in-plane electron-hole
correlations and fully includes the effect of the localization
potential. It obeys the two-dimensional $N$-particle  Bloch
equation which is obtained by averaging the three-dimensional
Bloch equation over $z$ and using Eq.~(\ref{3D_Ham}) and the
ansatz in Eq.~(\ref{adiab})
\begin{equation}
\frac{\partial}{\partial \beta}\, \rho({\bf R}^{xy},\beta) =
\left( - \sum\limits_{i=1}^{N_e,N_h} \frac{\hbar^2}{2 m_i}
\nabla^2_{xy} + V^{\, xy}_{\rm eff}+ V^{loc, \, xy}_{e(h)} \right)
\, \rho({\bf R}^{xy},\beta). \label{Ham_eff}
\end{equation}
Here, we have introduced an effective 2D in-plane interaction
potential $V_{\rm eff}^{\,xy}$:
\begin{eqnarray}
V_{\rm eff}^{\, xy}(\beta)= \int d Z_e\; d Z_h \;
\sum\limits_{i<j} \frac{e_i \, e_j}{\epsilon |{\bf r}_i - {\bf
r}_j|} \; \rho (Z_e,
\beta)\, \rho (Z_h, \beta) \nonumber \\
\times \left[\int d Z_e \; d Z_h
\; \rho (Z_e, \beta)\, \rho (Z_h, \beta)\right]^{-1}. \label{ueff}
\end{eqnarray}
and the total localization potential
\begin{eqnarray}
V^{loc, \, xy}_D =
\begin{cases}
E_0(L+\delta) - E_0(L), & \text{if $\sqrt{(x^2+y^2)} \leq D/2$}; \\
0, & \text{if $\sqrt{(x^2+y^2)} > D/2$}, \\
\end{cases}
\label{uloc2d}
\end{eqnarray}
where $E_0(L)$ is the lowest energy level in a
QW of widths $L$ (see inset of Fig.~1).

\section{Numerical Simulation Approach.}\label{PIMCA}
We numerically solve the Bloch equation~(\ref{Ham_eff}) using the
path integral representation of the density matrix. Using the
operator identity $e^{-\beta H}=( e^{-\tau H} )^M$, the density
matrix at inverse temperature $\beta$ can be expressed in terms of
$M$ density matrices, each taken at a $M$ times higher temperature
$M\, k_BT$, or as a path integral with $M$ steps of size $\tau =1/
(M k_BT)$~\cite{Feynm},
\begin{eqnarray}
\rho({\bf R}, {\bf R}; \beta)=\int d{\bf R}_1 \dots \int d{\bf
R}_{M-1} \, \sum_{P} \frac{(- 1)^{\delta P}}{N!} \nonumber
\\
\langle {\bf R} |e^{-\tau \hat H} | {\bf R}_{1} \rangle \dots
\langle {\bf R}_{M-1} |e^{-\tau \hat H} | {\hat P} {\bf R}
\rangle, \label{pimc}
\end{eqnarray}
where ${\bf R}$ represents a set of coordinates of $N$ particles
in $2$ dimensions; ${\hat
P}$ is the $N$-particle exchange operator, $(-1)^{\delta P}$
denotes the sign of the permutation for Fermi particles (electrons
and holes); $\rho({\bf R}, {\bf R}'; \tau) = \langle {\bf R}
|e^{-\tau \hat H} | {\bf R}' \rangle$ is the coordinate
representation of the $N$-particle density matrix at the new
inverse temperature $\tau$. For the $N$-particle high-temperature
density matrix, $\rho({\bf R}, {\bf R}'; \tau)$, we use the pair
approximation which is valid for $\tau \leq 1/(3 \,
H_a^{*})$,
\begin{eqnarray}
\rho({\bf R}, {\bf R}';\tau) \approx \prod_i^N \rho^{[1]} ({\bf
r}_i, {\bf r}_i'; \tau) \prod_{j<k} \frac{\rho^{[2]} ({\bf r}_j,
{\bf r}_k, {\bf r}_j', {\bf r}_k'; \tau)} {\rho^{[1]}({\bf r}_i,
{\bf r}_i'; \tau) \; \rho^{[1]} ({\bf r}_k, {\bf r}_k'; \tau)} +
O(\rho^{[3]}), \label{pair_dm}
\end{eqnarray}
where $i,\, j$ are particle indices and $\rho^{[1]} \; (\rho^{[2]})$ is the
one(two)-particle density matrix. The one-particle density matrix, $\rho^{[1]}$, is the
known free-particle kinetic energy density matrix. The pair density matrix $\rho^{[2]}$
was obtained from a direct numerical solution of the two-particle Bloch equation for
which we used the matrix squaring technique ~\cite{storer,ceperley95rmp}.

As one can see from Eq.~(\ref{pimc}), the needed diagonal matrix
elements of the low-temperature density operator are expressed in
terms of all diagonal and off-diagonal matrix elements of the
corresponding high-temperature density operator which can be
effectively computed using path integral Monte Carlo simulations,
see Ref.~\cite{png2} and references therein. Obviously, for these
simulations to be efficient, it is crucial that the off-diagonal
density matrix, $\rho^{[2]}$, can be quickly evaluated for any
given initial $(\textbf{r}_{i},\textbf{\, r}_{j})$ and final
$({\textbf{r}_{i}}',\,{\textbf{r}_{j}}')$ radius vectors of the
particle positions. For this reason, before doing the PIMC
simulations, we calculated in advance tables of the pair density
matrices (DM) for each type of interaction in our system. In our
electron-hole system in a QW with the localization potential, we
needed to calculate: i) three tables of pair density matrices
corresponding to electron-electron, hole-hole and electron-hole
interactions given by the two-particle Bloch equation with the
smoothened effective 2D Coulomb potential, see Eq.~(\ref{ueff}),
and ii) two tables with the density matrix of a single particle
(electron or hole) in a 2D cylinder of finite height (for
particles localized at the interface defect). The contributions of
all these interactions (correlations) can be treated as additive,
once the used high-temperature pair density matrices correspond to
sufficiently high temperature (such that commutators of pairs of
energy contributions are negligibly small). Finally, using the
pair DM tables, we are able to calculate the many-body density
matrix, Eq.~(\ref{pair_dm}), for  any set of initial~${\bf R}$ and
final~${\bf R}'$ positions of all particles. We substitute this
expression into Eq.~(\ref{pimc}) and perform the high dimensional
integration using a multilevel (bisection) Metropolis algorithm
(see, e.g.~\cite{ceperley95rmp}).

In the present calculations we used tables of the pair density
matrices at a temperature three times the effective electron-hole
Hartree, i.e. $1/\tau = 3 H_a^*= 403~K$. By choosing in
Eq.~(\ref{pimc}) the number of factors equal to $M=270$, the full
density matrix, $\rho({\bf R}, {\bf R}; \beta)$ and all
thermodynamic quantities can be accurately evaluated at a
temperature $T=1.49$~K. All results shown below correspond to this
temperature value.

Before considering in detail the effect of quantum well width
fluctuations on the binding energies of excitonic complexes, we
recall the main results obtained for {\em ideal} QW's with finite
width $L$ \cite{png2}. Quasi-two-dimensional systems like GaAs
QW's have been extensively investigated in the last years, both
experimentally~\cite{exp_xminus, exp_xplus, birkedal, Adachi,
Panthke, Bar-Ad, Kim, Smith, langbein} and
theoretically~\cite{Riva1,Riva2,Ess,Tsuch,Whitt}. These studies
revealed that, due to the confinement, the 2D excitonic states
have binding energies which are several times larger than the
binding energies in the bulk materials. This effect is mainly due
to the confinement of the carrier wave functions along the
structure growth direction, which leads to a two-dimensional
character of excitons and, consequently, to a change in the
in-plane interaction potential between the carriers. In the
framework of the {\em adiabatic approximation} these changes can
be easily seen in the effective in-plane potential $V_{\rm
eff}^{\,xy}$, Eq.~(\ref{ueff}),
\begin{figure}
\vspace{1.8cm} \hspace{-1.9cm}
\includegraphics[width=.50\textwidth]{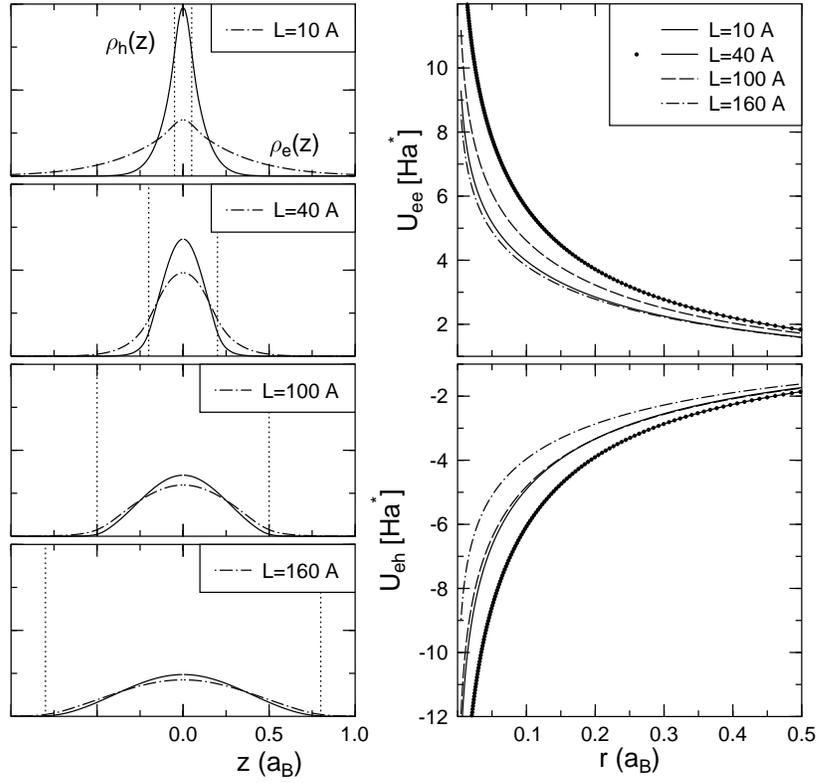}
\vspace{-0.4cm} \caption{{\em Left}: electron $\rho_e(z)$ and hole
$\rho_h(z)$ density matrix in the QW (dotted lines indicate QW
walls). {\em Right}: QW width dependence of the effective
electron-electron (ee) and electron-hole (eh) potentials, see
Eq.~(\ref{ueff}).} \label{dmatrix_ueff}
\end{figure}
which depends on the quantum well width $L$ as a result of the
integration in Eq.~(\ref{ueff}) over the free electron and hole
density matrices which reflect the probability distributions of an
electron and hole in the square well potential of given width $L$.
In Fig.~\ref{dmatrix_ueff}, we present the electron and hole
densities in the square well with a width varying between $10$
{\AA} and $160$ {\AA}. The results in the left panel of
Fig.~\ref{dmatrix_ueff} confirm that, due to the smaller mass, the
electron is less localized than the hole and, for $L \leq 20$
{\AA}, most of the electron density resides in the barrier
material.

The effective in-plane potential $V_{\rm eff}^{\, xy}$ is shown in
the right panel of Fig.~\ref{dmatrix_ueff}. Notice that it
depends on $L$ in a non-monotonic way reaching a maximum
(absolute) value around $L\approx 40$ {\AA}. Such an increase
with decreasing $L$ (for not too small values of $L$) is also
found experimentally and theoretically and is
due to an increase in the interparticle
correlations and results in the main contribution to the increase of
the binding energies in ideal quantum wells at intermediate QW
widths.

\section{Binding energies}\label{results1}
In this section we investigate the combined influence of the
finite QW width and of the interface defects (defect width and
height) on the ground state of the exciton and excitonic
complexes. In particular, we analyze the modification of the
binding energies and of the average interparticle distances in the
ground state of excitons (X), positive and negative trions
($X^{\pm}$) and biexcitons ($X_2$).

\subsection{Binding energies and average size of excitonic
complexes}
For an ideal QW, i.e. without interface defects, we  define the
binding energy of the exciton, charged exciton and biexciton as:
\begin{eqnarray}
E_B(X) &=& E_e + E_h - E(X), \nonumber
\\
E_B(X^{\pm}) &=& E(X) +E_{h(e)} - E(X^{\pm}), \nonumber
\\
 E_B(X_2) &=& 2 E(X) - E(X_2),
\label{exp_eb}
\end{eqnarray}
 where  $E_{e \, (h)}$ is the energy of a single electron
(hole) in the given quantum well with a free particle mean kinetic
(thermal) energy $k_B T$, and $E(A)$ is the total energy of the
excitonic complex $A$. If an interface defect is present and a
localization potential is included in our calculations, then the
above definitions must be modified. All energies must be replaced
by the corresponding energies of particles localized in the defect
potential. The corresponding generalized expressions will be given
in Sec.~\ref{loc_sec}.

Using a finite temperature approach such as PIMC, one calculates
states in thermal equilibrium. Moreover when the temperature
is not sufficiently low and comparable with the depth of the
trapping potential, the {\em equilibrium state} reached in a
sufficiently long simulation will correspond to non-localized
states rather than localized ones. To correctly obtain the total
and binding energies of localized excitonic complexes, the results
were computed not by averaging over all states, but by restricting
the average to the states localized in the trapping potential.

\begin{figure}
\includegraphics[width=.6\textwidth, angle=-90]{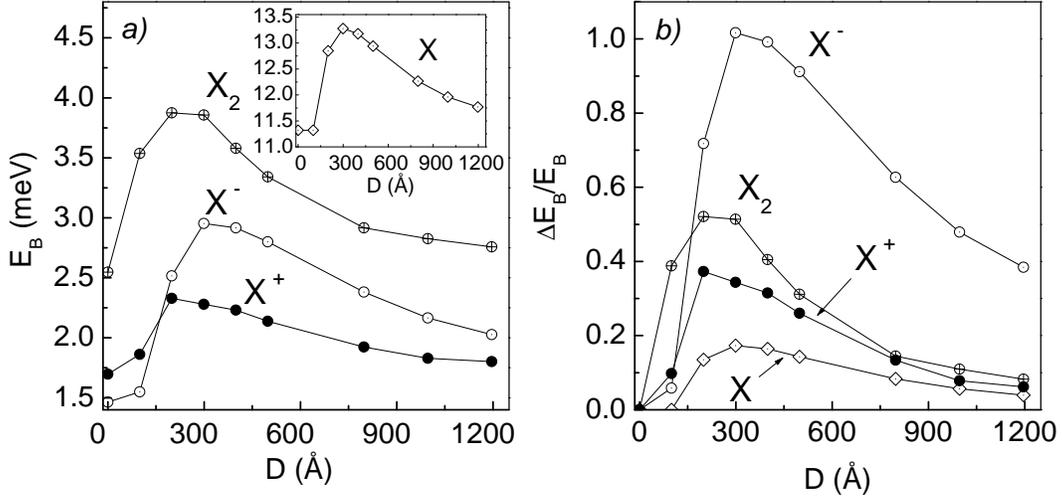}
\vspace{-2cm} \caption{(a): Binding energies of various excitonic
complexes vs. the diameter of the 1 ML quantum well width
fluctuation for a QW width of $L=60$ {\AA} and temperature,
$T=1.5$~K; (b): The same as (a) but now for the relative
increase of the binding energy.} \label{eb_locR}
\end{figure}

We now discuss the results for the binding energies and average
interparticle distance in the ground state of various excitonic
complexes as a function of the depth and width of the interface
defects. In Fig.~\ref{eb_locR}(a) we plot the binding energies
versus the diameter of the trapping potential, $D$, for the case
of a 1 monolayer (1 ML) surface defect. The corresponding relative
gain in the binding energies due to the interface defect is shown
in Fig.~\ref{eb_locR}(b). As an example, we took a QW width of $L
= 60$ {\AA} and a 1ML QW width fluctuation which corresponds to
the following heights of the electron and hole localization
potentials, $|V^{loc}_e|=3.43$ meV and $|V^{loc}_h|=1.28$ meV,
respectively. From
Fig.~\ref{eb_locR} one can notice that  for all excitonic
complexes the binding energy is always larger when a defect is
present than in the ideal case. In particular, it increases with
the diameter $D$ of the trapping potential up to some maximum
after which it slowly decreases. However, for very large $D$
the system approaches very slowly the ideal QW result, but 1 ML wider
than the original one.

Notice that the position of the maximum is different for the
different excitonic complexes. This is readily explained by the
different lateral size of different bound states which is
determined by the lateral extension of the electron and hole wave
functions in the trap and by their relative distance. The
electrons are more sensitive to the defect because the trapping
potential has a larger effect on their localization (the holes are
substantially localized even in the absence of the defect).
Further, we observe that the lateral confinement has a very
different effect on the magnitude of the exciton, trion and
biexciton binding energies, see Fig.~\ref{eb_locR}(b). In
particular, the exciton binding energy is only relatively weakly
affected by the localization, i.e. a very small peak in the
relative binding energy gain of less than $20 \%$ occurs. In
contrast, the binding energy of the negatively charged exciton
increases by more than $100 \%$, from $1.4$ meV to $3$ meV for the
localization potential of diameter $D \approx 300$ {\AA}.

\begin{figure}
\includegraphics[width=.6\textwidth, angle=-90]{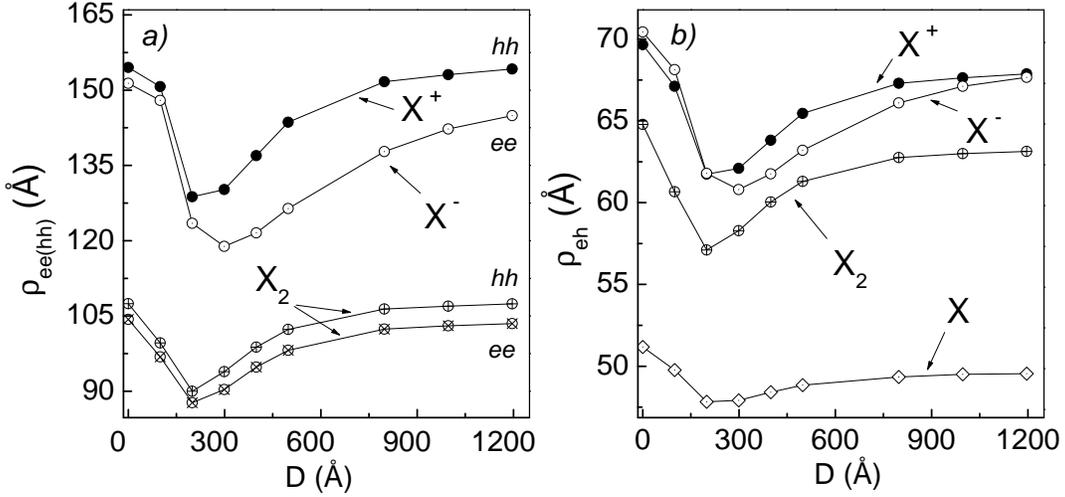}
\vspace{-2cm} \caption{Average distance between the constituents
of the different excitonic
complexes as a function of $D$ for (a) equally charged and (b)
oppositely charged particles. } \label{avrdist_locR}
\end{figure}

In Fig.~\ref{avrdist_locR} we show the average in-plane
interparticle distance, $\rho_{ij}$, versus the diameter of the
localization trap for the exciton, trions and biexciton.
For
a 2D system this results, after using the adiabatic approximation,
in the following expression:
\begin{equation}
\rho_{ij}= \int_0^{\infty} r_{ij} \; g(r_{ij}) \; d r_{ij} \bigg/
\int_0^{\infty} g(r_{ij}) \; d r_{ij}, \label{rho}
\end{equation}
where $g(r_{ij})$ is the pair distribution function of the
particles $i$ and $j$.

By comparing the electron-hole distances in various complexes, see
Fig.~\ref{avrdist_locR}(b), it can be seen that the electron-hole
distance in the exciton, i.e. the size of the exciton, is about
$1.2-1.4$ times smaller than the electron-hole separation in the
charged excitons and about $3$ times smaller than the average
electron-electron (hole-hole) distances,
Fig.~\ref{avrdist_locR}(a). This explains our previous finding,
see Fig.~\ref{eb_locR}(b), that the exciton state is much less
influenced by the lateral confinement than the $X^-$. In the
exciton, where the electron and hole are coupled much stronger,
the interparticle distance changes only slightly with the diameter
$D$ and the effect on their binding energy is weak. Notice also
that the peak in the gain of the binding energy quite closely
follows the minima of the electron-hole interparticle distances.
This result agrees well with the experimental findings
(discussed below) that in the case of localized particles the
binding energy of the $X^{-}$ exceeds the one of the $X^{+}$.

Further, it is interesting to note that the biexciton appears to
be {\it less extended} than the trions, thus explaining the fact
that trions have a lower binding energy than the biexciton, see
Fig.~\ref{eb_locR}(a). This is consistent with the experimental
observations, which will be discussed in detail in
Sec.~\ref{experiment}. At the same time, the biexciton is more
affected by the interface defect than the positive trion, see
Fig.~\ref{eb_locR}(b). This suggests that the number of electrons
in the excitonic complexes plays a much more important role in the
interaction with the interface defect than the number of holes. In
fact, both, the $X^-$ and $X_2$, which contain two electrons, are
more influenced by the localization than the $X$ and $X^+$. The
reason is that the localization potential has a stronger impact on
the confinement of electrons than on holes, as noted above.

Next we compare the negative and the positive trion. For small
localization islands, $D \leq 150$ {\AA}, the average distances
between electrons in $X^-$ and holes in $X^+$ are very similar
and, consequently, the relative gain in the binding energies of
the two trions is close as well, see Fig.~\ref{eb_locR}(b). In
contrast, for wide localization islands, i.e. $D \geq 200 $ {\AA},
the behavior of the two trions differs significantly, e.g. the
binding energies and the interparticle distances between the
respective pair of equally charged particles deviate from each
other, see Fig.~\ref{eb_locR} and Fig.~\ref{avrdist_locR}. The
reason is a different influence of the QW potential on the
electron and hole wave functions in the $X^{-}$ and $X^{+}$
states. In a quantum well the electrons are substantially extended
into the barrier material (i.e. in the $z$-direction), whereas the
holes are much more confined, see Fig.~\ref{dmatrix_ueff}(a). This
leads to a significantly larger overlap of the two electrons in
the $X^-$ state compared to the two holes in the $X^{+}$ and to a
weaker in-plane effective electron-electron interaction potential,
Eq.~(\ref{ueff}). As a result, in Fig.~\ref{avrdist_locR}(a) at
$D=0$ we can see that the electron-electron distances,
$\rho_{ee}$, are slightly smaller than those of the holes,
$\rho_{hh}$. Though the average distances in the $X^-$ and $X^+$
states are very similar, if one compares fluctuations of the
average distances, $\delta \rho =\sqrt{\langle \rho^2\rangle -
\langle \rho\rangle^2}$, they will be much larger for the
electrons in the negative trion $X^-$. This can be seen directly
from the behavior of the pair distribution functions shown in
Figs.~\ref{pair_dist}(a,b). When the trions are localized by the
additional lateral confinement these fluctuations are quenched and the
distance between particles of the same charge is decreased, e.g.
for $0 \leq D \leq 300$ {\AA}. This must have a stronger effect on
the electrons than on the holes. The stronger repulsive
interaction between the holes in the $X^+$ state reduces to a
larger extent the gain in the binding energy as compared to the
$X^-$ state, where the electron interaction is weaker and
particles can be brought to smaller distances. This accounts for
the increase of the $X^-$ binding energy up to, and even beyond, the
$X^+$ binding energy for large $D$.

\subsection{Origin of the enhanced binding energies}\label{loc_sec}
It turns out that the differences in the trion binding energies
are not caused solely by their different spatial extensions, see
Fig.~\ref{avrdist_locR}. In fact, the difference between the
average interparticle distances in the two trions is not
sufficiently large to account for the gain of the binding energy
for the $X^-$ and the $X^+$, see Fig.~\ref{eb_locR}(b). The
explanation must be found in the fact that the gain in the binding
energy, as a function of the size of the localization island, does
not come only from the changes in the Coulomb interaction related
to the interparticle distances, but also from changes (and
differences) of the electron and hole localization energies, and
the kinetic energy of the particles.

In the presence of the localization potential each electron and
hole acquires an additional (single particle) potential energy -
the localization energy $E_{loc}^{e(h)}$. This single particle
energy can be
directly computed by averaging the localization potential over the
radial electron (hole) distribution,
\begin{equation}
E_{loc}^{e(h)}(D)=\int dr_{e(h)}\, g^R(r_{e(h)})\,
V^{loc}_D(r_{e(h)})/\int dr_{e(h)} \, g^R(r_{e(h)}). \label{Eloc}
\end{equation}
Similarly, each bound excitonic complex is affected by the
localization potential where each electron and hole contributes
additionally to the total localization energy
\begin{equation}
E_{loc}(X)=E_{loc}^e(X)+E_{loc}^h(X), \label{eloc_x}
\end{equation}
where $E_{loc}^{e(h)}(X)$ for the exciton is computed in an
analogous way as was done for the single particles in
Eq.~(\ref{Eloc}), but with the appropriate radial electron (hole)
distribution inside the localized exciton. Further,
Eq.~(\ref{Eloc}) can be straightforwardly generalized to the
trions and biexciton cases.

Obviously, the localization energy modifies the total energy of
all particles,
\begin{eqnarray}
E(X)=E(X,D=0) \rightarrow E(X,D); \; E(X^{\pm})\rightarrow
E(X^{\pm},D); \; E(X_2)\rightarrow E(X_2,D),
\end{eqnarray}
where for all bound states ($X,X^{\pm},X_2$) the total energy can
now be written as
\begin{equation}
E(D)=E(0)+\delta E_{\text{Coul}}(D) + \delta E_{\text{kin}}(D) +
E_{loc}(D). \label{eq_c}
\end{equation}
Here $\delta E_{\text{Coul}}(D)$ and $\delta E_{\text{kin}}(D)$
denote, respectively, the change of the Coulomb and kinetic energy
due to the presence of the defect of diameter $D$. From
Eqs.~(\ref{exp_eb}) and~(\ref{eq_c}) we can now derive the
definition of the binding energy of an exciton in the presence of
a localization potential:
\begin{equation}
E_B(X,D)=E_B(X,0)+\delta E_{\text{Coul}}(X,D) + \delta
E_{\text{kin}}(X,D) + \delta E_{loc}(X,D), \label{eq_eb1}
\end{equation}
where we define the change of the localization energy due to the
excitonic bound state
\begin{eqnarray}
\delta E_{loc}(X,D) &=& \delta E_{loc}^e(X,D) + \delta
E_{loc}^h(X,D), \nonumber
\\
\delta E_{loc}^{e(h)}(X,D) &=& E_{loc}^{e(h)}(D)-
E_{loc}^{e(h)}(X,D)=\int dr_{e(h)}\, [ g^{R}(r_{e(h)}) -
g^{R}(r_{e(h)},X)]\, V^{loc}_D \label{eq_eb2}.
\end{eqnarray}
Similarly, the change of the kinetic and interaction energies of
an electron-hole pair forming an exciton is given by
\begin{eqnarray}
\delta E_{\text{kin}}(X,D) &=& E_{\text{kin}}^e(D) +
E_{\text{kin}}^h(D) - E_{\text{kin}}(X,D) \nonumber
\\
\delta E_{\text{Coul}}(X,D) &=& - E_{\text{Coul}}^{eh}(X,D).
\label{eq_eb3}
\end{eqnarray}
These expressions can be generalized directly to the case of the
trions and biexciton. For example, for the positive trion the
change in the Coulomb interaction is expressed as follows:
\begin{equation}
\delta E_{\text{Coul}}(X^+,D) = \{E_{\text{Coul}}^{eh}(X)\} -
\{2E_{\text{Coul}}^{eh}(X^+) + E_{\text{Coul}}^{hh}(X^+) \}.
\label{eq_eb4}
\end{equation}

In Eqs.~(\ref{eq_eb3})-(\ref{eq_eb4}) the Coulomb energy,
$E_{\text{Coul}}(D)$, is estimated as an average of the effective
potential, $V_{\rm eff}^{\, xy}$ (Fig.~\ref{dmatrix_ueff}), over
the pair distribution functions calculated for each type of
interparticle interaction. For example, for the electron-hole
interaction in the exciton we have:
\begin{equation}
E^{eh}_{\text{Coul}}(X)=\int dr \, V_{eh}^{\, xy}(r) \,
g_{eh}(r,X).
\end{equation}
The kinetic energy of the localized single electron (hole),
$E^{e(h)}_{\text{kin}}$, the localized exciton,
$E_{\text{kin}}(X)$, the trion, $E_{\text{kin}}(X^+)$, and the
biexciton, $E_{\text{kin}}(X_2)$, were computed as the difference
between the total energy and the full potential energy which
includes both Coulomb interaction and localization energy,
$E_{\text{kin}}=E - (E_{\text{Coul}}+ E_{loc})$. This result can
be compared with a more strict thermodynamic estimator of the
kinetic energy as the mass derivative of the partition function,
$E_{\text{kin}}=\frac{m}{\beta Z}\frac{\partial Z}{\partial m}$.
We found that both expressions give very similar results.

\begin{figure}
\hspace{-1cm}
\includegraphics[width=.7\textwidth, angle=-90]{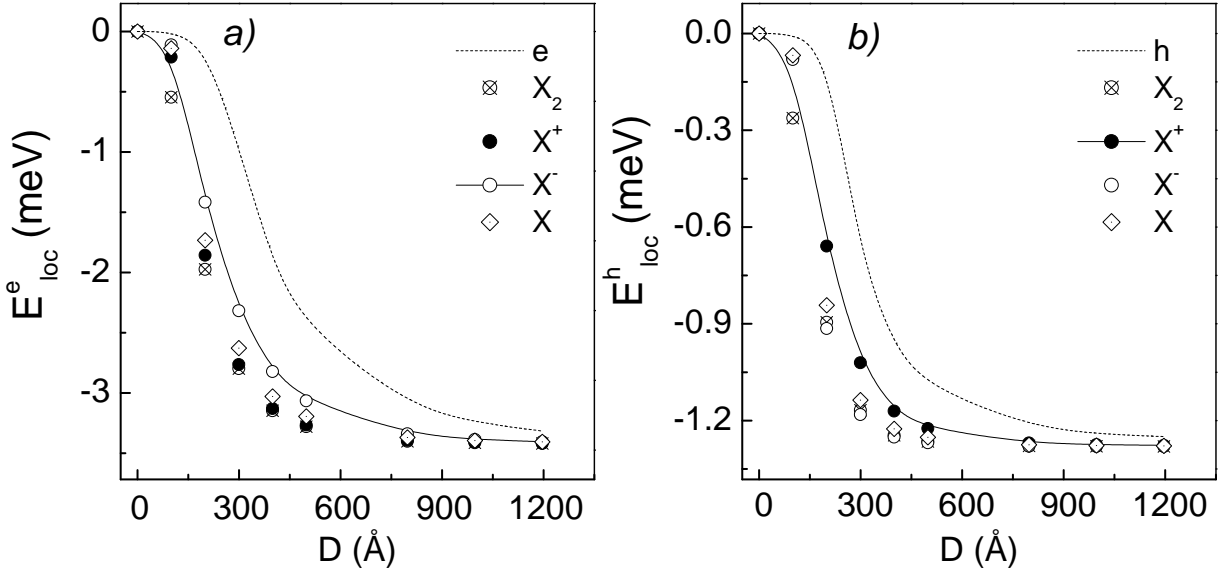}
\vspace{-3cm} \caption{Localization energy of an electron (a) and
hole (b) in the different excitonic complexes as a function of
diameter of the localization island for a one monolayer
fluctuation and QW width of $L=60$\AA. Dotted lines are the
corresponding localization energies of a single electron and hole
in the same localization potential.} \label{loc_U}
\end{figure}

In Fig.~\ref{loc_U} we show the contribution of the well width fluctuation,
which we will call the localization energies, to the single
(non-interacting) particles, $E^{e(h)}_{loc}$, and interacting
electrons (holes) in various excitonic complexes:
$E^{e(h)}_{loc}(X), E^{e(h)}_{loc}(X^{\pm}), E^{e(h)}_{loc}(X_2)$.
This energy was estimated as an average of the
localization potential, $V^{loc, \, xy}_D$ (Eq.~(\ref{uloc2d})),
over the radial distribution functions calculated separately for
electrons and holes, see e.g. Figs.~\ref{pair_dist}~(c,d).
For large D, $E_{loc}^{e(h)}$ should approach the depth of the
in-plane localization potential, i.e. $|V^{loc}_e|=3.43$meV and
$|V^{loc}_h|=1.28$meV.
As one
can see the interparticle interaction significantly increases the
localization. Both, in the left and right panels of
Fig.~\ref{loc_U}, the localization energy of a single particle
(electron or hole) shown by the dotted line is much less (in
absolute value) compared to that of the $X, X^{\pm}, X_2$ and
saturates only for $D \geq 1200$ {\AA}. The explanation is that
the attraction between electrons and holes already leads to
significant spatial localization of the particles compared to the
free particle thermal wavelength and, as a consequence, the
effective localization potential felt by each particle (the
potential is smoothened over the particle's wave functions) is
deeper.

An interesting point that we notice from Fig.~\ref{loc_U} is that,
on average, the electrons and holes in the biexciton are more
localized than in all other bound states. The only exception is
the localization energy of the hole in the range $200$ {\AA} $\leq
D \leq 300$ {\AA} when a single hole in the $X^-$ state tends to
be more localized, see Fig.~\ref{loc_U}(b). Among all considered
bound states only the biexciton appears to be strongly localized
for islands with a diameter around $D\approx 100$ {\AA} (other
excitonic states, in our simulations at temperature $T\approx
1.5$~K, have a much higher probability to become delocalized due
to thermal fluctuations). This is confirmed by the gain in the
binding energy shown in Fig.~\ref{eb_locR}(b), where at the point
$D=100$ {\AA} only the biexciton shows a strong increase by $40
\%$, and in Fig.~\ref{loc_U} at $D=100$ {\AA} only the biexciton
shows nonnegligible values for the localization energy,
$E^e_{loc}=0.54$ meV and $E^h_{loc}=0.25$ meV, for the electron
and the hole respectively.

Now, using the values of the localization energies of the
electrons and the holes shown in Fig.~\ref{loc_U}, one can
estimate the total localization energies of the different
excitonic complexes. E.g. the localization energies of the
negative trion and the biexciton can be estimated as follows:
\begin{equation}
E_{loc}(X^-)=2E^e_{loc}(X^-)+E^h_{loc}(X^-), \quad
E_{loc}(X_2)=2E^e_{loc}(X_2)+2E^h_{loc}(X_2) \label{loc_eng}.
\end{equation}
These energies include many-body correlation effects in
combination with the specific radial distribution functions which
are different for each bound state, see e.g.
Figs.~\ref{pair_dist}~(c,d).

\begin{figure}
\hspace{-1cm}
\includegraphics[width=.55\textwidth, angle=-90]{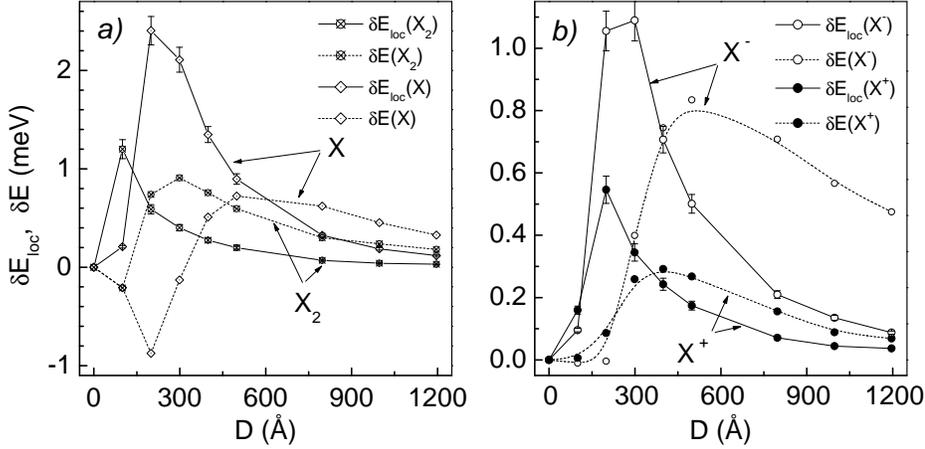}
\vspace{-2cm} \caption{(a) Localization energy gain, $\delta
E_{loc}$~(Eq. (\ref{eq_eb2})), and energy gain, $\delta E$, of the
exciton and biexciton vs. diameter of the localization island. (b)
Comparison of the same energies for positive and negative trions.
The same parameters are used as in Fig.~5.} \label{loc_U2}
\end{figure}

\begin{figure}
\hspace{-1cm}
\includegraphics[width=.55\textwidth, angle=-90]{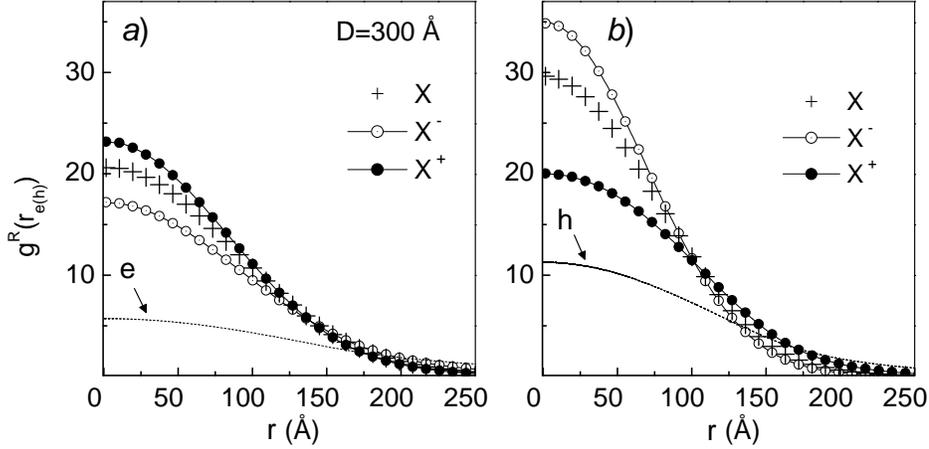}
\vspace{-2cm} \caption{(a,b) Radial distribution function of
electrons (a) and holes (b) for exciton, positive and negative
trions, and the single electron (hole) as a function of distance,
$r_{e(h)}$, from the center of the localization potential. The
same parameters are used as in Fig.~5.} \label{pair_dist}
\end{figure}

We now analyze the binding energies which are modified from their
ideal expression, Eq.~(\ref{exp_eb}), due to the three localization
corrections, cf. Eqs.~(\ref{eq_eb1},\ref{eq_eb2}), namely the
contributions due to changes of the Coulomb interaction, kinetic
and localization energy. Each part can be calculated separately
and the results are presented in Figs.~\ref{loc_U2}(a,b). Solid
lines in this figure show changes of the localization energy,
$\delta E_{loc}(D)$, of the exction, trions, biexciton as a
function of the diameter of the island, $D$. Dotted lines show
combined changes in the Coulomb interaction and kinetic energy
minus the binding energy in the same QW without the interface
defect, $\delta E(D)= \delta E_{\text{Coul}}(D) + \delta
E_{\text{kin}}(D) - E_B(0)$. Consequently,
\begin{equation} E_B(D)=E_B(0) + \delta E_{loc}(D)+\delta E(D).
\label{new_eb}
\end{equation}

From Fig.~\ref{loc_U2}(a) we note that the difference in the
localization energies, shown by the solid curves in
Figs.~\ref{loc_U2}(a,b), is maximal for the exciton. This is easy
to understand because for the exciton (see Eq.~(\ref{eq_eb2})) we
subtract from the energies of the unbound electron and hole (which
are less localized) the localization energy of a more localized
bound electron-hole pair in the exciton state (see
Fig.~\ref{loc_U}). According to Eq.~(\ref{new_eb}) this gives
positive contributions to the binding energy, e.g. at the point
$D=300$ {\AA} it is about $2$ meV. For the biexciton and the negative
trion the difference of localization energies reaches a maximum
value of about $1$ meV. For the biexciton the maximum is reached
around the defect diameter $D\approx 100$ {\AA}, for $X^-$ around
$D\approx 300$ {\AA}, and for $X^+$ at $D\approx 200$ {\AA}.

The dotted lines in Figs.~\ref{loc_U2}(a,b) are the contributions
due to the defect induced changes of the Coulomb interaction and
kinetic energy. In this way the total binding
energy~(\ref{new_eb}) is the sum of the respective solid and
dotted curves in Fig.~\ref{loc_U2}. By comparing each pair of
curves for $X$, $X^{\pm}$ and $X_2$, one can note a general
feature valid for all bound states: at $D \leq 300$ {\AA} the main
contribution to the binding energy comes from changes of the
localization energy, Eq.~(\ref{eq_eb2}), but when the defect
diameter exceeds $D\approx 600$ {\AA} the main effect is due to
changes of the kinetic energy and the interparticle interaction.

It is interesting that for the exciton and biexciton the quantity
$\Delta E$ becomes negative in the range of defect diameters $D
\leq 300$ {\AA} which leads to a reduction of the binding energy.
The reason is that $\delta E$ is composed of $\delta
E_{\text{kin}}$ which is negative and the positive term $\delta
E_{\text{Coul}}$. $\delta E_{\text{kin}}$ is negative because the
kinetic energies of bound particles are larger than those of
unbound particles. This is readily understood because the wave
functions of bound particles are more localized (spatially less
extended) and thus have a larger curvature, which increases the
kinetic energy.
It follows also from the virial theorem that the
increase of potential (or interaction) energy leads to an increase
of the kinetic energy.

The comparison of $\delta E_{loc}$ and $\delta E$ between the
positive and negative trions in Fig.~\ref{loc_U2}(b) shows that
the negative trion is more affected by the localization for $D
\geq 100$ {\AA}. This has a direct relation with
Fig.~\ref{eb_locR}, where the $X^-$ binding energy exceeds the one
of the $X^+$ for all $D \geq 150$ {\AA}.


The positive trion $X^+$ is a much heavier composite particle with
two holes and one electron compared to the $X^-$, where there are
two electrons and one hole. As a consequence the $X^+$ is less
mobile and, thus, is less affected by the lateral confinement.
This fact can be demonstrated by Figs.~\ref{pair_dist}(a,b) where
we show the radial distribution functions $g^R(r_{e(h)})$ of an
exciton, negative/positive trions and a single electron/hole in a
QW of width $L=60$ {\AA} for a defect diameter $D=300$ {\AA}. The
panels (a),(b) show the radial density of electrons and holes,
respectively. Notice that at the center of the localization
confinement ($r_{e(h)}=0$) the radial density of electrons in the
$X^-$ state is increased by more than a factor of $3$ as compared
to the radial density of a single electron in the same
localization potential (curve indicated by ``e''). At the same
time, comparison of the densities for the holes in the $X^+$ state
and the density of the single hole shows only an increase by a
factor of $1.8$. These results are in agreement with our previous
statement that the localization has a larger effect on electrons
than on holes and, consequently, on the composite particles with a
larger number of electrons (in our case $X^-$ and $X_2$). From
Figs.~\ref{pair_dist}(a,b) we can also see that the central radial
density of electrons is highest in the $X^+$ state, whereas the
highest central radial density of holes is observed in the $X^-$.
In both panels (a) and (b) the electron and hole radial densities
near the localization potential center for the exciton state lies
in between the corresponding values for the $X^+$ and $X^-$. This
fact is easily understood from the symmetry of the spatial
configuration of particles in the $X$, $X^+$ and $X^-$ states in
the cylindrical defect potential. For trions, the single hole in
the $X^-$ (the single electron in the $X^+$) will most probably
occupy the central position in between the two electrons (two
holes) to minimize the correlation energy and thus the total
energy.

\subsection{Dependence on the number of monolayer fluctuations}
\begin{figure}
\includegraphics[width=.55\textwidth, angle=-90]{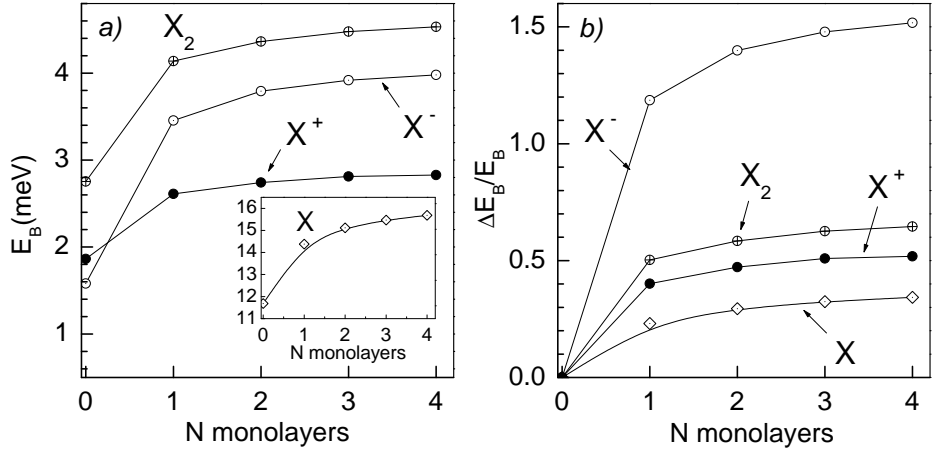}
\vspace{-2cm} \caption{The same as Fig.~\ref{eb_locR}, but now as
a function of the number of monolayers, $N$, for a fixed QW width
of $L=40$ {\AA}, temperature $T=1.5$~K and diameter of the
localization potential, $D=400$ {\AA}.} \label{eb_locU}
\end{figure}

Now we allow for well width fluctuations larger than 1 ML and
analyze the dependence of the binding energies and interparticle
distances on the number of monolayers. We fix the quantum well
width to $L=40$ {\AA} and the diameter of the defect to $D=400 $
{\AA}. In Fig.~\ref{eb_locU} we plot the binding energy of the
different excitonic complexes as a function of the number of
monolayers $N$ forming the defect (the curves are guides to the
eye since $N$ is a discrete index). Notice that increasing $N$
leads to a monotonic increase in the binding energy for all
exciton complexes which saturates for $N\approx 4$. Notice that
the increase of the binding
energy of the biexciton and of the negative trion is almost
parallel. This
is a clear confirmation of our earlier conclusion
that the lateral confinement of the electron by the defect has a more
pronounced effect on the binding energy than the hole
confinement. Again we observe that the binding energy of the $X^-$
exceeds that of the $X^+$ in the presence of a localization
potential (see the discussion of Fig.~\ref{eb_locR}).
Fig.~\ref{eb_locU} shows that this trend persists in the case of
increasing defect depth.

\begin{figure}
\includegraphics[width=.55\textwidth, angle=-90]{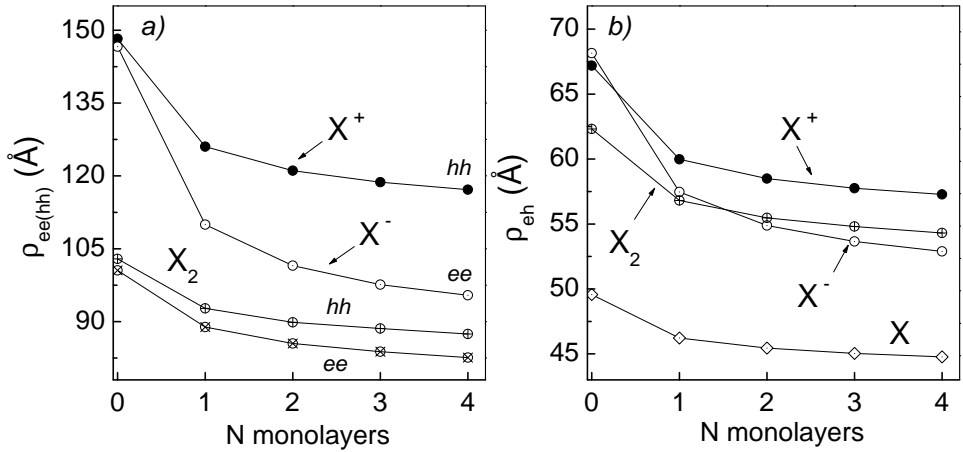}
\vspace{-2cm} \caption{The same as Fig.~\ref{avrdist_locR}, but
now as a function of the number of monolayers, $N$. Same
parameters are used as in Fig.~\ref{eb_locU}.} \label{avrdist_locU}
\end{figure}

Fig.~\ref{avrdist_locU} displays the dependence of the mean e-e,
h-h and e-h distances in the different excitonic states on the
number of monolayers. All distances (i.e. the spatial extension of
all bound states) decrease monotonically with $N$ and saturate
around $N=4$.

\section{Comparison with experiment}\label{experiment}

Before making a comparison between our theoretical results and the
experimental data it is necessary to point out a problem that may
arise in such a comparison. In particular, one has to be aware of
the fact that for finite temperatures there exists an experimental
uncertainty about the state in which particles remain after
recombination. Namely, particles produced after recombination will
typically have a finite kinetic energy which may allow them to
leave the localization potential. In this case, the energy of the
emitted photon will be reduced by an amount needed to overcome
the height of the localization potential. As a consequence, the
photoluminescence lines of the exciton complexes exhibit an
additional broadening due to the finite kinetic energy of the
remaining particles, thus making the determination of the binding
energy more complicated.

Secondly, in experiments there may be two types of measurements of
excitons and trions. One, when the excitonic complexes are probed
in a single well. These measurements may favor observation of the
most strongly localized excitons and trions (as it was found in
Ref.~\cite{Allan}), while measurements from a QW ensemble favor
higher-lying states nearer to the continuum. In the latter case
the localization effects does not strongly influence the observed
ensemble QW spectra. As the theory shows, this would affect the
binding energy, as the localization effects are of great
importance in narrow QW's.

\subsection{Exciton binding energy}

\begin{figure}[h]
\hspace{-1cm}
\includegraphics[width=.6\textwidth, angle=-90]{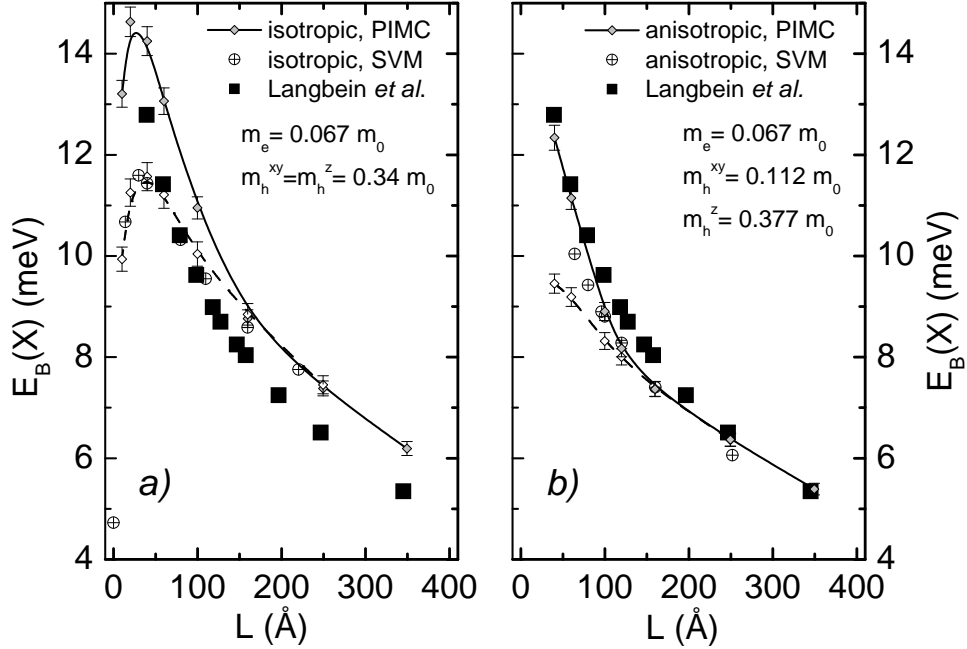}
\vspace{-0cm} \caption{Exciton binding energies for isotropic (a)
and anisotropic (b) hole mass vs. quantum well width.
Solid square symbols are experimental
data of Ref.~\cite{langbein} and circular symbols with a plus
sign are obtained from the stochastic variational
calculations~\cite{Riva1,Riva2}. Full (dashed) lines with small
symbols are PIMC results when the localization is included (not
included).} \label{ani_Ex}
\end{figure}

We first compare in Figs.~\ref{ani_Ex}(a,b) the theoretical
and experimental binding
energies of the exciton as function of the QW width.
Solid and dashed curves in these figures show the binding energy
in the QW with defect and in the ideal QW without interface
roughness, respectively. The localization potential is considered
as due to a well width fluctuation of one monolayer over a circular
area of diameter $D=400$ {\AA} in accordance with
the experimental findings of Ref.~\cite{Gam}. As we
can see from Fig.~\ref{eb_locR} for the QW width around $60$ {\AA}
this gives an upper bound to the localization effect on the
binding energies. In order to be consistent we used the same
localization potential when calculating the binding energies of
trions and the biexciton.

When using the isotropic approximation for the hole mass ($m_e =
0.067 \, m_0$, $m_h^{xy}= m_h^z= 0.34 \, m_0$), we notice from
Fig.~\ref{ani_Ex}(a) that the theoretical results systematically
overestimates the binding energy. For example, for $L \geq 100$
{\AA} the theory gives binding energies which are approximately
$10 \%$ larger than the experimental results. In order to see the
effect of the anisotropy of the hole band we performed the
calculation using an anisotropic hole mass ($m_h^{xy}=0.112 m_0$
in the QW plane and $m_h^z=0.377 m_0$ in the QW growth
direction~\cite{Winkler}) which brings the theoretical points
significantly closer to the experiment, see Fig.~\ref{ani_Ex}(b).
For the QW widths $L \leq 100$ {\AA} good agreement is found when
the localization effects are included in our model. For example,
in Ref.~\cite{Brunner} for a QW width $L=34$~{\AA} a value
$12.4$~meV has been measured for a localized exciton. This agrees
quite well with our theoretical prediction of about $12.3$~meV
for a comparable QW structure ($L=40$~{\AA}).

However, there is still a slight discrepancy between theory and
experiment for QW widths in the range $ 100 \leq L \leq 200$
{\AA}. In this range, in our theoretical model, the localization
effects are negligible for excitons. For $L \geq 150$ {\AA} the
two curves calculated with and without localization practically
coincide. In this case, for the wide QW's ($L \geq 150$ {\AA}) the
theory appears to agree quite well with the experimental binding
energies. This allows us to conclude that for such quantum well
structures the localization does not play significant role for
excitons and they are not trapped by the interface defects.

In Figs.~\ref{ani_Ex}(a,b) we compare our results with those
obtained with the stochastic variational
approach~\cite{Riva1,Riva2}, both for the isotropic and
anisotropic hole masses which gives additional credit to the
accuracy of our numerical approach. Note that the hole mass and
the dielectric constant used in Ref.~\cite{Riva1,Riva2} ($m_e =
0.067 \, m_0, m_h^{xy}= 0.099 \, m_0, \epsilon=12.1$) were
slightly different from ours. This leads to minor discrepancies
for the binding energies for QW widths smaller than $100$ {\AA}
because the electron and hole densities in the growth direction
are very sensitive to the QW confinement and to the chosen values
of electron/ hole masses in narrow QW's (see
Fig.~\ref{dmatrix_ueff}).

The PIMC results (for the anisotropic hole mass and the 1 ML
interface defect of diameter $D=400$ {\AA}) can be compared with
the those of Refs.~\cite{Bastard,Heller}, where the {\em exciton
binding energy to the interface defect}, $E^D(X)$ has been
calculated. This quantity is defined as the difference of the total
exciton energy with and without the defect potential. In
particular, the variational calculations of Ref.~\cite{Heller} for
two types of exciton trial wavefunctions give, for a QW of width
$L=35$ {\AA} and $D=400$ {\AA}, $E^D(X)=7.4$~meV and $10.4$~meV,
respectively (the isotropic electron mass is
$m_e^z=m_e^{xy}=0.0782 m_0$, the heavy hole masses are the same as
in the present work). The PIMC calculations for the QW width
$L=40$ {\AA} give a similar value, $E^D(X)=8.16$ meV. In the
variational calculations of Ref.~\cite{Bastard} a value
$E^D(X)=3.4$ meV has been reported for a QW with $L=70$ {\AA}. The
present calculations give $E^D(X)=3.72$ meV and $E^D(X)=0.84$ meV,
for a QW width $L=60$ {\AA} and $L=100$ {\AA} respectively.
This trend definitely shows that the excitons become less
localized with increasing QW width.

In conclusion, the comparison of Figs.~\ref{ani_Ex}(a) and (b)
shows that taking into account the anisotropy of the hole mass
leads to a decrease of the exciton binding energy compared to the
isotropic case. For example, in a $50$ {\AA} wide QW this amounts
to about $2$ meV, and in a $250$ {\AA} wide QW it is about $1$
meV.

\subsection{Binding energy of positive and negative trions}

\begin{figure}
\hspace{-1cm}
\includegraphics[width=.6\textwidth, angle=-90]{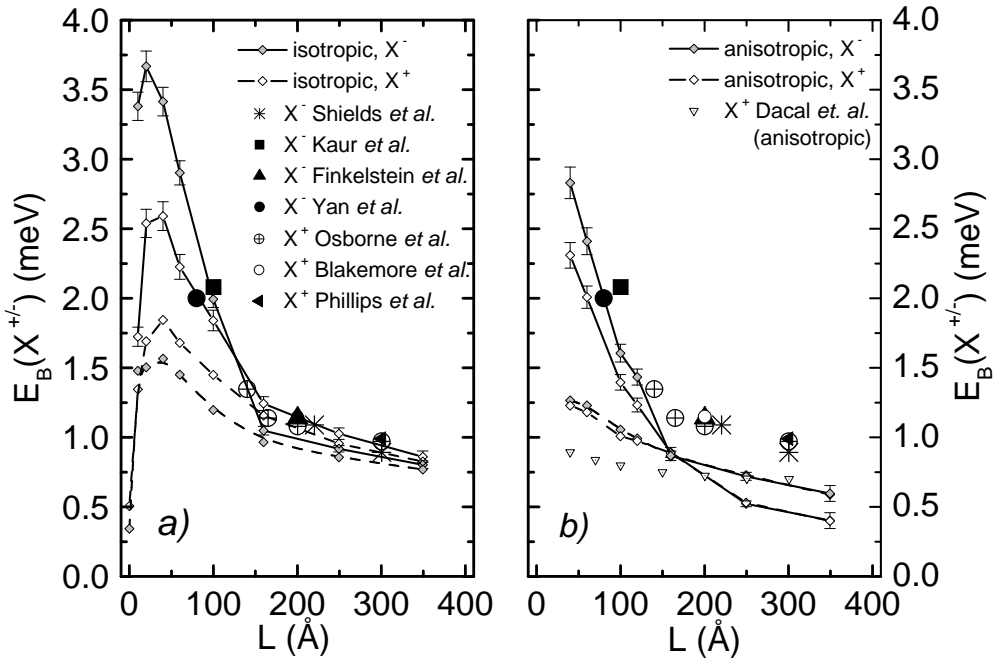}
\vspace{-0cm} \caption{Trion binding energies for isotropic (a)
and anisotropic (b) hole mass vs. quantum well width. Symbols are
experimental data of Refs.~\cite{exp_xminus, exp_xplus} and
theoretical calculations of Ref.~\cite{Dacal}. Full (dashed) lines
with symbols are PIMC results when localization is included (not
included).} \label{eb_trions}
\end{figure}

In Figs.~\ref{eb_trions}(a,b) we present our results for the
binding energy of the trions. We compare our theoretical results
with the available experimental data for
negative~\cite{exp_xminus} and positive~\cite{exp_xplus} trions,
and variational calculations of Ref.~\cite{Dacal}. First, we can
note that the agreement with the experiments is quite good for QW
widths $L \geq 150$ {\AA}. Specifically, the experimental points
for the $X^+$~\cite{exp_xplus} are close to our theoretical
result, see Fig.~\ref{eb_trions}(a). Unfortunately, for narrow
QW's ($L \leq 150$ {\AA}) when localization effects become
important, there are currently no available experimental data.
This would be of high interest as the two points for the $X^-$
reported by R.~Kaur {\em et al.} and Z.C.~Yan {\em et
al.}~\cite{exp_xminus} show a more rapid increase of the binding
energy with the QW width than the one predicted by theory when the
localization is not taken into account. On the contrary,
calculations with the QW width fluctuations included agree well
with these data.

Further, in Fig.~\ref{eb_trions}(a) we observe a crossing of the
two binding energy curves ($X^+$ and $X^-$) with the localization
included near the point $L=135$ {\AA}, and for narrow wells the
binding energy of the negative trion becomes larger. For example,
in a $40$ {\AA} quantum well $E_B(X^-) \approx 3.4$ meV while for
the positive trion $E_B(X^+) \approx 2.6$ meV. Preliminary
experimental investigations of the $X^-$ and $X^+$ in the presence
of localization~\cite{Allan} seem to confirm these findings and
give for a $50$ {\AA} wide QW $E_B(X^-)=3.27$ meV and
$E_B(X^+)=2.35$ meV. It is interesting to notice that in an ideal
QW the situation is the opposite and the binding energy of the
$X^+$ is always larger than the one of the $X^-$. As was already
discussed in section~\ref{loc_sec} the reason is that the
localization has a stronger influence on the electrons and,
consequently, on the negative trion.

When the anisotropy of the hole mass is included in our
calculations we found that there is no crossing between the
binding energy of the $X^+$ and the $X^-$. A comparison with the
variational calculations of L.~Dacal {\em et al.}~\cite{Dacal}
(see Fig.~\ref{eb_trions}(b)) also done with the anisotropic hole
mass, but slightly different parameters $V_e (V_h)=224.5$ meV
($149.6$ meV) and $\epsilon=13.2$ (compared to ours $V_e
(V_h)=216$ meV ($163$ meV) and $\epsilon=12.58$), shows some
discrepancy with the present calculations for $X^+$ binding energy
in the narrow QW's. This can be due to the following reason.
Variational calculations strongly depend on a form of used trial
wave functions. In particular, the Eq.~(3) in Ref.~\cite{Dacal}
becomes less accurate in the narrow quantum wells and hence
requires a larger set of variational parameters. This could lead
to a better agreement with the PIMC results.

In Ref.~\cite{Dacal2} the effect of localization (at the interface
defect with a cylindrical symmetry and a Gaussian shape) on the
trion binding energy was considered for the anisotropic hole mass
with the same parameters as in Ref.~\cite{Dacal}. It was found
that, with the defect present, the $X^-$ binding energy is increased
from $0.4$ to $0.6$ meV in the $150$ {\AA} wide QW. However, these
values are much lower than the PIMC results which show an
increase from $0.90$ to $1.02$ meV. The results of Ref.~\cite{Dacal2}
are even lower than the value $E_B(X^-)=0.75$ meV reported in
Ref.~\cite{Dacal} for the same QW width but without the
localization effect included. In Ref.~\cite{Dacal2} the number of
basic variational trial wavefunctions was reduced compared to
Ref.~\cite{Dacal} and, as the above comparison shows, this appears to
be not sufficient for a quantitative description of the trion.

Other theoretical calculations done with an isotropic hole mass
and in the absence of a localization potential shown in
Fig.~\ref{eb_trions}(a) agree quite well with our data for the
ideal QW case. For example, both in Ref.~\cite{riva9} and
Ref.~\cite{Tsuch} it was found that the $X^+$ binding energy is
larger by about $20 \%$ than the $X^-$ binding energy. This is in
agreement with most experimental results which show that the $X^+$
has a binding energy which is larger than or close to the one of
the $X^-$~\cite{exp_xminus}. However, the theoretical results of
Ref.~\cite{Stebe2000} for a $300$ {\AA} wide QW showed that
$E_B(X^+)$ is lower than $E_B(X^-)$ which is opposite to the
results of Refs.~\cite{Tsuch,riva9}, but the latter is in
agreement with our results for the case of the anisotropic hole
mass (see Fig. 11(b)). From the other hand, the anisotropic
calculations show not satisfactory agreement between theory and
experiment, in particular for $L>150$ {\AA} where theoretical
curves are about $0.5$ meV below the experimental points. Here the
localization effects (in the framework of our model) are not
important and can not be the reason for this disagreement.

\subsection{Binding energy of biexciton and the Haynes factor}

\begin{figure}
\hspace{-1cm}
\includegraphics[width=.6\textwidth, angle=-90]{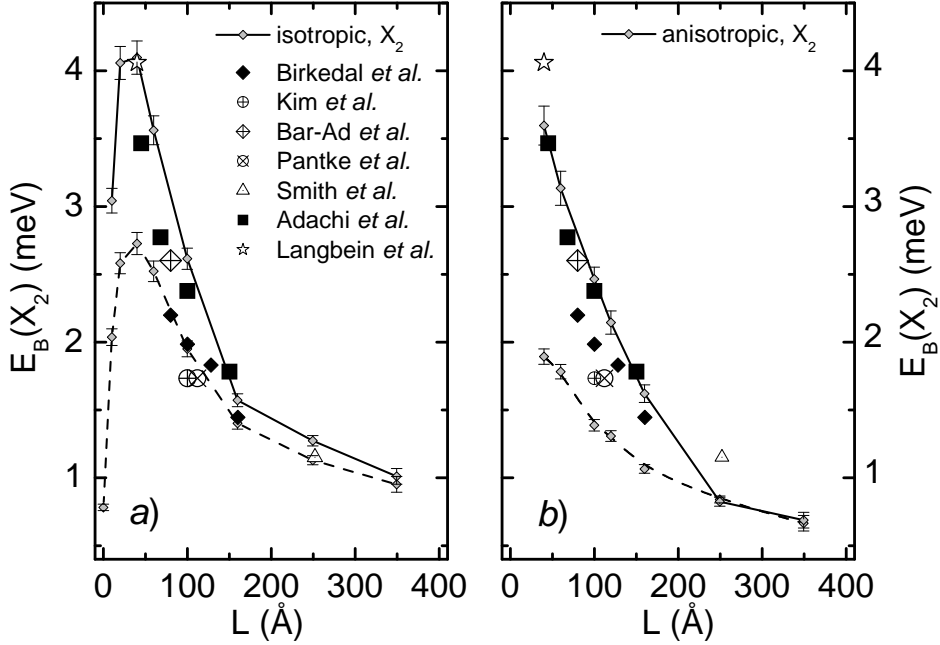}
\vspace{-0cm} \caption{The same as in Fig.~\ref{eb_trions} but now
for the biexciton binding energy. Symbols are experimental data of
Refs.~\cite{birkedal, Adachi, Panthke, Bar-Ad, Kim,
Smith,langbein}.} \label{eb_biexc}
\end{figure}

Now we consider the biexciton binding energy. Solid (dashed)
curves in Figs.~\ref{eb_biexc}(a,b) show our theoretical results in
the presence (without) of well width fluctuations which we compare
with available
experiments~\cite{birkedal, Adachi, Panthke, Bar-Ad, Kim, Smith,
langbein} (symbols in figure).
In the case of an isotropic hole mass, we found that with
localization taken into account (full line in
Fig.~\ref{eb_biexc}(a)) the theoretical curve passes through the
data of Refs.~\cite{Adachi, Bar-Ad}, but the experimental data of
Refs.~\cite{birkedal, Panthke, Kim, Smith} agrees with the theory for
an ideal QW (see the dashed curve in Fig.~\ref{eb_biexc}(a)). The
inclusion of the localization effects brings the theoretical curve slightly
above the experiment, predicting that in e.g. a narrow $40$ {\AA} wide
QW the binding energy of the biexciton is about $(4.0-4.1)$ meV. It is
interesting that practically the same experimental result is
reported in Ref.~\cite{langbein}, where the value $4.1$ meV is
found for the $40$ {\AA} QW. It should be stressed however, that
our results for the binding energy, in the presence of a one
monolayer well width fluctuation, are close to an upper limit,
since the used defect diameter ($D=400$ {\AA}) was such that
it gave practically the
maximal gain in the biexciton binding energy as due to
localization (see Fig.~\ref{eb_locR}). But this value of the
defect diameter, $D=400$ {\AA} was found to be a very good
estimate of the characteristic defect size in GaAs quantum
wells~\cite{Gam}.

The use of an anisotropic hole mass (see Fig.~\ref{eb_biexc}(b))
leads to a reduction of the biexciton binding energy by almost
$(0.4 - 0.8)$ meV. As in the case of trions, for wide QW's with $L
\geq 150$ {\AA} the anisotropic approximation gives a less
satisfactory agreement with the experimental points. From the
other hand, for the QW widths $L \leq 150$ {\AA} we found
marvellous agreement with the experiment of Ref.~\cite{langbein}
when localization is included in our calculations (see
Fig.~\ref{eb_biexc}(b), solid curve).

For an anisotropic hole mass and a defect depth of 1 ML,
we can compare our results with the variational calculations of
Ref.~\cite{Heller}. In these calculations, however, a repulsion
between particles of the same charge was not taken into account,
which makes the comparison only qualitative. In Ref.~\cite{Heller}
the biexciton binding energy of about $1.6$ meV and $1.3$ meV have
been reported for the $30$ {\AA} and $50$ {\AA} wide QW's, respectively,
and a defect diameter $200$ {\AA}. These values of the binding energies
are less than our result, $1.9$ meV, for a non-localized biexciton
in a $50$ wide {\AA} QW. With the interface defect ($D=400$
{\AA}), the PIMC calculations show an increase of the binding
energy up to $3.6$ meV, which is close to the experimental data.
For example, in Ref.~\cite{Brunner}, a value $E_B(X_2)=4.2$ meV was
attributed to a localized biexciton in a $34$ {\AA} wide QW.

In conclusion, the present comparison of the theory and the
experiment allows us to conclude that the anisotropic
approximation for the hole mass gives better agreement for
narrow QW's (with L < 150 {\AA}). On the other hand, for wide
QW's ($L > 150$ {\AA}) our model calculations show that the use of
isotropic approximation show better agreement. We expect
that the accuracy of the calculations (and of the anisotropic
approximation, in particular) can be further improved by taking
into account the mismatch of dielectric constants and
particle masses in the well and barrier materials. This should lead
to a better agreement with the experiment.

In comparing the theory and the experiment one should keep in mind
that different experimental results have been obtained from
different quantum wells which have not been grown under the same
conditions and, consequently, their well width fluctuations may
also be different. Nevertheless, in overall, the present
calculations show that even a simple model of localization can
satisfactory explain the experimental data on the binding
energies. We expect that our theory can make reliable predictions
if detailed information on the quantum well fluctuations in the
samples would be available.

\begin{figure}
\hspace{-1cm}
\includegraphics[width=.6\textwidth, angle=-90]{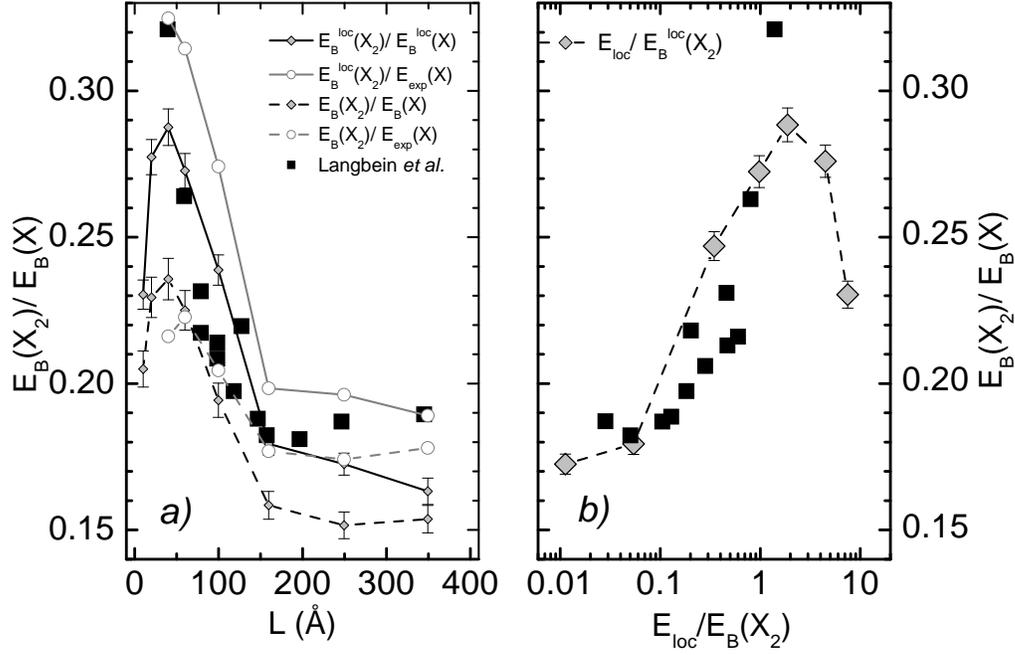}
\vspace{-0cm} \caption{The biexciton to exciton binding energy ratio
(Haynes factor), (a) versus QW width and (b) vs. normalized
localization potential. Symbols: experimental data for
GaAs/Al$_{0.3}$Ga$_{0.7}$As QWs \cite{langbein}. Curves with
symbols: PIMC results. Solid (dashed) curve is the exciton binding
energy with (without) localization included. (b) $E_B(X_2)/E_B(X)$
versus localization strength $E_{\text{loc}}/E_B(X_2)$. Squares
are the experimental data. Rhombus are PIMC results.}
\label{EX_2_Ex}
\end{figure}

Next, we consider the so-called {\em Haynes factor}, which is the
ratio between the biexciton and the exciton binding energies,
$\nu=E_B(X_2)/E_B(X)$. We compare our results, see
Fig.~\ref{EX_2_Ex}(a), with the experimental data of
Ref.~\cite{langbein} (solid squares), where the influence of
localization on the binding energies of the exciton and biexciton
was considered. Here all theoretical curves are given only for the
isotropic hole mass ($m_h^{xy}=m_h^z=0.34$) and, as
in previous figures, the solid (dashed) line is for the case
(without) localization included. The ratio of the exciton and
biexciton energies calculated using an anisotropic hole mass is
very similar and is therefore not shown.

The lines with small filled symbols are obtained by using
the theoretical values of the exciton and biexciton binding
energies (Figs.~\ref{ani_Ex}(a) and ~\ref{eb_biexc}(a)), which are
denoted as $E_B^{\text{loc}}(X_2)/E_B^{\text{loc}}(X)$ and
$E_B(X_2)/E_B(X)$. In addition, since there is a discrepancy
between the experimental and the theoretical results in Fig.~\ref{ani_Ex}(a)
for the exciton binding energy which affects the Haynes
factor, we calculate also the Haynes factor as the ratio of the
theoretical biexciton binding energy to the exciton energy from
the experiment~\cite{langbein} (shown in Fig.~\ref{ani_Ex}(a) by the
solid squares). The corresponding results are denoted in
Fig.~\ref{EX_2_Ex}(a) as $E_B^{\text{loc}}(X_2)/E_{\text{exp}}(X)$
and $E_B(X_2)/E_{\text{exp}}(X)$.

First, we note that the Haynes factor is practically independent of the
QW width for $L\geq 150$ {\AA} in agreement with
experiment. However, the value of the
constant is different in various cases. When localization is included
(solid line) our calculations show a systematic increase of the
Haynes factor $\nu=E_B(X_2)/E_B(X)$ by up to $17\%$. For the exciton
and biexciton localized on the interface defect the Haynes factor
is $\nu\approx 0.175$, while for the ideal QW, $\nu\approx 0.15$.
For well widths $L \ge 50$ {\AA}, all theoretical and experimental
results exhibit a monotonic decrease of the Haynes factor with
increasing $L$. One can observe that the experimental results are
located mainly between the two theoretical curves corresponding
to the localized and not localized biexciton. Most of the points lie
on the dashed curve with the open circles,
$E_B(X_2)/E_{\text{exp}}(X)$, suggesting that the experimentally
measured binding energies correspond to non-localized biexcitons,
but some of the data are substantially above the dashed curve and
agree better with the assumption of predominantly localized
biexcitons. This is the case for well width $L \leq 80$ {\AA}
where a strong increase of the Haynes factor is found also from
the theory. For example, our theory gives a maximum value $\nu = 0.24$
without localization and $\nu=0.29$ for localized particles.

Notice that the agreement between the experimental and theoretical
results with localization is also confirmed by
Fig.~\ref{EX_2_Ex}(b), where the Haynes factor is plotted against
the normalized localization potential. We compute the localization
energy $E_{loc}$ as the difference between the energy of the
localized and non-localized excitonic complex (in
Ref.~\cite{langbein} the localization energy was defined from the
full width at half maximum of the heavy-hole exciton absorption
line). As follows from our theory, below $150$ {\AA} well
thickness when the localization energy becomes of the order of the
biexciton binding energy an enhancement of the Haynes factor is
observed. This indicates that localization has a crucial effect on
the Haynes factor and it must be taken into account for a correct
interpretation of the experimental results.

\section{Conclusion}\label{conclusion}
In summary, using a path integral Monte Carlo approach, we made
a detailed analysis of different excitonic complexes in $Ga As/Al
Ga As$ quantum wells. We calculated and analyzed the exciton,
trion and biexciton binding energies, pair distribution functions
and mean interparticle distances in the excitonic complexes in a
wide range of QW widths. Our method is based on first principles
and does not invoke expansions in eigenfunctions. The approach is
general, flexible and -- what is also important -- is not limited
to certain specific symmetries of the particle wave function. It
allows for a simultaneous account of QW confinement, localization
and valence band anisotropy, and thus can give an accurate
theoretical treatment of many experimental systems. The only
assumption -- the adiabatic approximation (Eq.~(\ref{adiab})) --
appears to be justified for the present application.  Simple
estimates show that its accuracy may be reduced for wide QW's
with $L \geq 250$ {\AA}.

Extending our previous analysis \cite{png2}, we concentrated here
on the influence of disorder, i.e. of the effect of QW width fluctuations on the
binding energies. The observed increase of the binding energies in
the presence of disorder is in good quantitative agreement with
the available experimental data. Furthermore, we analyzed the
influence of valence band anisotropy (hole mass) and found that,
in some cases (in particular for the exciton and biexciton binding energy) this
effect is important in order to achieve agreement with the
experimental data.

We also analyzed the case of deep interface defects corresponding
to several monolayers depth and found that this can give an
additional $20-30 \%$ increase of the binding energy compared to
the single monolayer case. This increase is even more pronounced for
the negative trion and the biexciton.

The present analysis is the first one in which exciton, trions and biexciton
are treated on an equal footing and in which the same size and
shape of the QW width fluctuation is invoked for different QW widths.
We assumed a 1ML QW width fluctuation over a circular area of diameter
$D=400$\AA. No other fitting parameters were introduced. This lead to
an overall good agreement of the well width dependence of the exciton,
trions and biexciton binding energies. It is expected that a better fit
with experiment is possible if, e.g. we allow for a non circular
shape of the well width fluctuation where the anisotropic localization
potential will be related to specific crystallographic directions.

Our results have two important implications which can be useful in
the interpretation of experimental data. First, by comparing
the measured binding energy with our numerical calculations for different
defect sizes allows one to characterize certain experimental
parameters, such as the magnitude of the disorder in a given
sample. Secondly, one can verify or predict whether or not the observed
excitonic
states are localized or delocalized in a
given experimental set up.

\section{Acknowledgement}
The authors would like to thank A.S.~Barker and D.~Gammon for
making available their experimental results~\cite{Allan} prior
to publication. This work was supported by the Flemish Science
Foundation (FWO-Vl), the Belgian Interuniversity Attraction
Poles (IUAP), the European Commission GROWTH programme NANOMAT project,
contract No. G5RD-CT-2001-00545 and by grants for CPU time at the Rostock
Linux-Cluster ``Fermion''. A.~Filinov gratefully acknowledges
hospitality of the Physics Department of the University of
Antwerp. Clara Riva is a FWO-Vl Postdoctoral Fellow.


\begin{thebibliography}{99}

\bibitem{exp_xminus}
R.~Kaur, A.J.~Shields, J.L.~Osborne, M.Y.~Simmons, D.A.~Ritchie,
and M.~Pepper, Phys. Stat. Sol. (a) {\bf 178}, 465 (2000);
G.~Finkelstein, H.~Shtrikman, I.~Bar-Joseph, Phys. Rev. B {\bf
53}, R1709 (1996); A.J.~Shields, M.~Pepper, D.A.~Ritchie,
M.Y.~Simmons, and G.A.C.~Jones, Phys. Rev. B {\bf 51}, 18\, 049
(1995); Z.C.~Yan, E.~Goovaerts, C.~Van~Hoof, A.~Bouwen, and
G.~Borghs, Phys. Rev. B {\bf 52}, 5907 (1995).

\bibitem{exp_xplus}
J.L.~Osborne, A.J.~Shields, M.~Pepper, F.M.~Bolton, and
D.A.~Ritchie, Phys. Rev. B {\bf 53}, 13002 (1996); J.S.~Blakemore,
J. Appl. Phys. {\bf 53}, R123 (1982); J.C.~Phillips, {\em Bonds
and Bands in Semiconductors}, (Academic Press, London, 1973).

\bibitem{Brunner} K.~Brunner, G.~Abstreiter, G.~B\"ohm,
G.~Tr\"ankle, and G.~Weimann, Phys. Rev. Lett. {\bf 73}, 1138
(1994).

\bibitem{birkedal}
D.~Birkedal, J.~Singh, V.G.~Lyssenko, J.~Erland, and J.M.~Hvam,
Phys. Rev. Lett. {\bf 76}, 672 (1996).

\bibitem{Adachi}
S.~Adachi, T.~Miyashita, S.~Takeyama, and Y.~Takagi, A.~Tackeuchi,
M.~Nakayama, Phys. Rev. B {\bf 55}, 1654 (1997).

\bibitem{Panthke}
K.H.~Pantke, D.~Oberhauser, V.G.~Lyssenko, J.M.~Hvam, and
G.~Weimann, Phys. Rev. B {\bf 47}, 2413 (1993).

\bibitem{Bar-Ad}
S.~Bar-Ad and I.~Bar-Joseph, Phys. Rev. Lett. {\bf 68}, 349
(1992).

\bibitem{Kim}
J.~Kim, D.~Wake, and J.~Wolfe, Phys. Rev. B {\bf 50}, 15099
(1994).

\bibitem{Smith}
G.~Smith, E.~Mayer, J.~Kuhl, and K.~Ploog, Solid State Commun.
{\bf 92}, 325 (1994).

\bibitem{langbein}
W.~Langbein and J.M.~Hvam, Phys. Rev. B {\bf 59}, 15 405 (1999).

\bibitem{review} F.M.~Peeters, C.~Riva, and K.~Varga, Physica B
{\bf 300}, 139 (2001).

\bibitem{Whitt} D.M.~Whittaker and A.J.~Shields, Phys. Rev. B {\bf 56}, 15185
(1997).

\bibitem{Riva1} C.~Riva, F.M.~Peeters, and K.~Varga, Phys. Rev. B {\bf 63},
115302 (2001).

\bibitem{Riva2} C.~Riva, F.M.~Peeters, and K.~Varga, Phys. Rev. B {\bf 61},
13873 (2000).

\bibitem{Ess}
A.~Esser, E.~Runge, R.~Zimmermann, and W.~Langbein, Phys. Rev. B
{\bf 62}, 8232 (2000).

\bibitem{Tsuch}
T.~Tsuchiya and S.~Katayama, {\em Proceedings of the 24th
International Conference on The Physics of Semiconductors,
Jerusalem, 1998}, edited by D.~Greshoni (World Scientific,
Singapore, 1999).

\bibitem{Stebe2000}
B.~St\'eb\'e and A.~Moradi, Phys. Rev. B {\bf 61}, 2888 (2000).

\bibitem{Eytan} G.~Eytan, Y.~Yayon, M.~Rappaport, H.~Shtrikman,
and I.~Bar-Joseph, Phys. Rev. Lett. {\bf 81}, 1666 (1998).

\bibitem{Gam} J.G.~Tischler, A.S.~Bracker, D.~Gammon, and D.~Park, Phys. Rev. B {\bf 66},
081310(R) (2002).

\bibitem{Bastard} G.~Bastard, C.~Delalande, M.H.~Meynadier,
P.M.~Frijlink, and M.~Voos, Phys. Rev. B {\bf 29}, 7042 (1984).

\bibitem{Heller} O.~Heller, Ph.~Lelong, and G.~Bastard, Phys. Rev.
B {\bf 56}, 4702 (1997).

\bibitem{Dacal} Luis C.O.~Dacal, R.~Ferreira, G.~Bastard, and Jos\'e
A.~Brum, Phys. Rev. B {\bf 65}, 115324 (2002).

\bibitem{Dacal2} Luis C.O.~Dacal, R.~Ferreira, G.~Bastard, and Jos\'e
A.~Brum, Phys. Rev. B {\bf 65}, 115325 (2002).

\bibitem{png2}
A.~Filinov, M.~Bonitz, and Yu.E.~Lozovik, phys. stat. sol. (c)
{\bf 238}, 1441 (2003) and in: {\em Progress in Nonequilibrium
Green´s Functions II}, M.~Bonitz and D.~Semkat (eds.) (World
Scientific, Singapore 2003), p. 436.

\bibitem{Feynm} R.P.~Feynman and A.R.~Hibbs, {\em Quantum Mechanics and Path Integrals}
(McGraw Hill, New York, 1965).


\bibitem{Winkler}
R.~Winkler, Phys. Rev. B {\bf 51}, 14 395 (1995).

\bibitem{storer}R.G.~Storer, J. Math. Phys. \textbf{9}, 964 (1968); A.D.~Klemm,
and R.G.~Storer, Aust. J. Phys. \textbf{26}, 43 (1973).

\bibitem{ceperley95rmp}D.M.~Ceperley, Rev. Mod. Phys. \textbf{65}, 279
(1995).

\bibitem{Allan}
A.S.~Bracker, D.~Gammon et al., to be published (private
communication).

\bibitem{riva9}
C. Riva, F.M. Peeters, and K. Varga, Phys. Rev. B {\bf 64}, 235301 (2001).

\end{thebibliography}
\end{document}